\documentclass[10pt,twocolumn,oneside,conference]{IEEEtran}
\usepackage{cite}
\usepackage{graphicx}
\usepackage{amsmath}
\usepackage{times}
\usepackage{latexsym}
\usepackage{graphicx}
\usepackage{bm}
\usepackage{amssymb}
\usepackage[center]{caption2}
\usepackage{stfloats}
\usepackage{cases}
\usepackage{array}
\usepackage{setspace}
\usepackage{fancyhdr}
\usepackage{algorithm}
\usepackage{algorithmic}

\usepackage{amsmath,amsthm}
\usepackage{graphicx}
\usepackage{subfigure}

\usepackage{xcolor}

\usepackage{balance} 
\usepackage{flushend} 

\allowdisplaybreaks[4]

\newcaptionstyle{mystyle1}{%
  \centering TABLE  \captiontext \par}
\captionstyle{mystyle1}
\newcaptionstyle{mystyle2}{%
  \captionlabel $.$ \: \captiontext \par}
\captionstyle{mystyle2}
\newcaptionstyle{mystyle3}{%
  \captionlabel $.$ \: \captiontext \par}
\captionstyle{mystyle3}

\begin{document}

\title{A CMDP-based Approach for Energy Efficient Power Allocation in Massive MIMO Systems}



\author{
\IEEEauthorblockN{Peng Li$^\dag$, Yanxiang Jiang$^{\dag*}$, Wei Li$^\ddag$, Fuchun Zheng$^\S$, and Xiaohu You$^\dag$}
\IEEEauthorblockA{$^\dag$National Mobile Communications Research Laboratory,
Southeast University, Nanjing 210096, China.\\
$^\ddag$Dept. of Information and Communication Engineering,
Xi'an Jiaotong University, Xi'an 710049, China.\\
$^\S$School of Systems Engineering, University of
Reading, Reading, RG6 6AY, UK.\\
$^*$E-mail: yxjiang@seu.edu.cn
}}


\maketitle

\begin{abstract}

    In this paper, energy efficient power allocation for the uplink of a multi-cell massive MIMO system is investigated.
    With the simplified power consumption model, the problem of power allocation is formulated as a constrained Markov decision process (CMDP) framework with infinite-horizon expected discounted total reward, which takes into account different quality of service (QoS) requirements for each user terminal (UT).
    We propose an offline solution containing the value iteration and Q-learning algorithms, which can obtain the global optimum power allocation policy.
    Simulation results show that our proposed policy performs very close to the ergodic optimal policy.

\end{abstract}

%
%

\section{Introduction}

    With the rapid development of wireless communication system, there has been a new surge of interest in energy efficient systems, due to the contradiction between the ever-increasing energy demand and the societal and economical concerns.
    As one of key technologies of 5G mobile communication systems, massive MIMO has been put forward to significantly improve the system capacity with extra degrees of freedom which facilitate transmit diversity and spatial multiplexing gains \cite{Noncooperative}.

    Recently, there has been an increasing research interest in energy efficiency (EE) for massive MIMO systems.
    As discussed in \cite{Optimal}, it is of primary importance to set up an accurate power consumption model for reliable guidelines of EE optimization.
    By using a refined power consumption model, closed-form EE-optimal value of transmit power was derived in \cite{Optimal} by means of some properties of Lambert \textit{W} function.
    However, the optimization problem there without any constraints on quality of service (QoS) failed to model the real scenario in communication systems.
    In the uplink of massive MIMO systems, the maximum transmit power and the minimum data rate for each user terminal (UT) should be included into basic QoS requirements.
    In \cite{Tradeoff}, the problem of maximizing the EE as a function of the numbers of UTs and antennas in BS was analyzed, for a given spectral efficiency and fixed transceiver power consumption parameters.
    Similarly, the impact of system parameters (the average channel gain to the UTs and the power consumption parameters) on the optimal EE was studied in \cite{Impact} for maximizing the EE with a fixed sum spectral efficiency.
    Besides the theoretical analysis on the relationships between system parameters and the optimal EE, it is of great importance to develop optimization methods for maximizing EE under the multi-cell scenario.

    More recently, the Markov decision process (MDP) method has been utilized to deal with the resource allocation problems for communication systems.
    In \cite{SMDP}, by using the semi-MDP method, a resource allocation scheme was proposed to achieve the optimal power efficiency for QoS-guaranteed services in OFDMA multi-cell cooperation networks.
    However, the technicalities and complexities associated with semi-MDP seldom lead to practical algorithms \cite{MIMO}.
    On the other hand, in order to meet the QoS requirements, only a few works on the constrained Markov decision process (CMDP) method for resource allocation in MIMO systems \cite{MIMO} and OFDM systems \cite{OFDM} have been reported. The problem of power and rate allocation in MIMO systems was modeled as a CMDP in \cite{MIMO} with the goal of minimizing the transmit power subject to delay constraints, while the problem of power and subcarrier allocation for downlink OFDMA systems was formulated as a CMDP in \cite{OFDM} with the goal of maximizing the EE under average delay constraints. By introducing a middle state called ``post-decision state", an online solution was proposed in \cite{OFDM}.

    Motivated by the aforementioned results, we propose a novel offline power allocation scheme to achieve the global optimum EE under QoS constraints in the uplink of multi-cell massive MIMO systems, which exploits the powerful optimization tool, constrained Markov decision process (CMDP).
    The power allocation policy is determined via the use of value iteration and Q-learning algorithms.
    The appeal of the value iteration algorithm is attributed to its ease in implementation and simplicity in the convergence condition to the global optimum solution.
    More importantly, the value iteration algorithm can be used for further studies to analyze the structure of the optimal policy obtained in this paper.
    The global convergence of the Q-learning algorithm guarantees the proposed offline solution to obtain the global optimum power allocation policy.
    Specifically, the proposed offline solution can exploit the obtained decision rule to build an offline look-up table, which can avoid the frequent and continuous computations and provide flexibility by adjusting the corresponding parameters of the value iteration and Q-learning algorithms.

    The rest of this paper is organized as follows.
    In Section II, the system model is briefly described.
    The problem formulation and solution algorithm are presented in Section III.
    Simulation results are shown in Section IV.
    Final conclusions are drawn in Section V.


\section{System Model}

\begin{figure}[!t]
\centering
\includegraphics[width=0.48\textwidth]{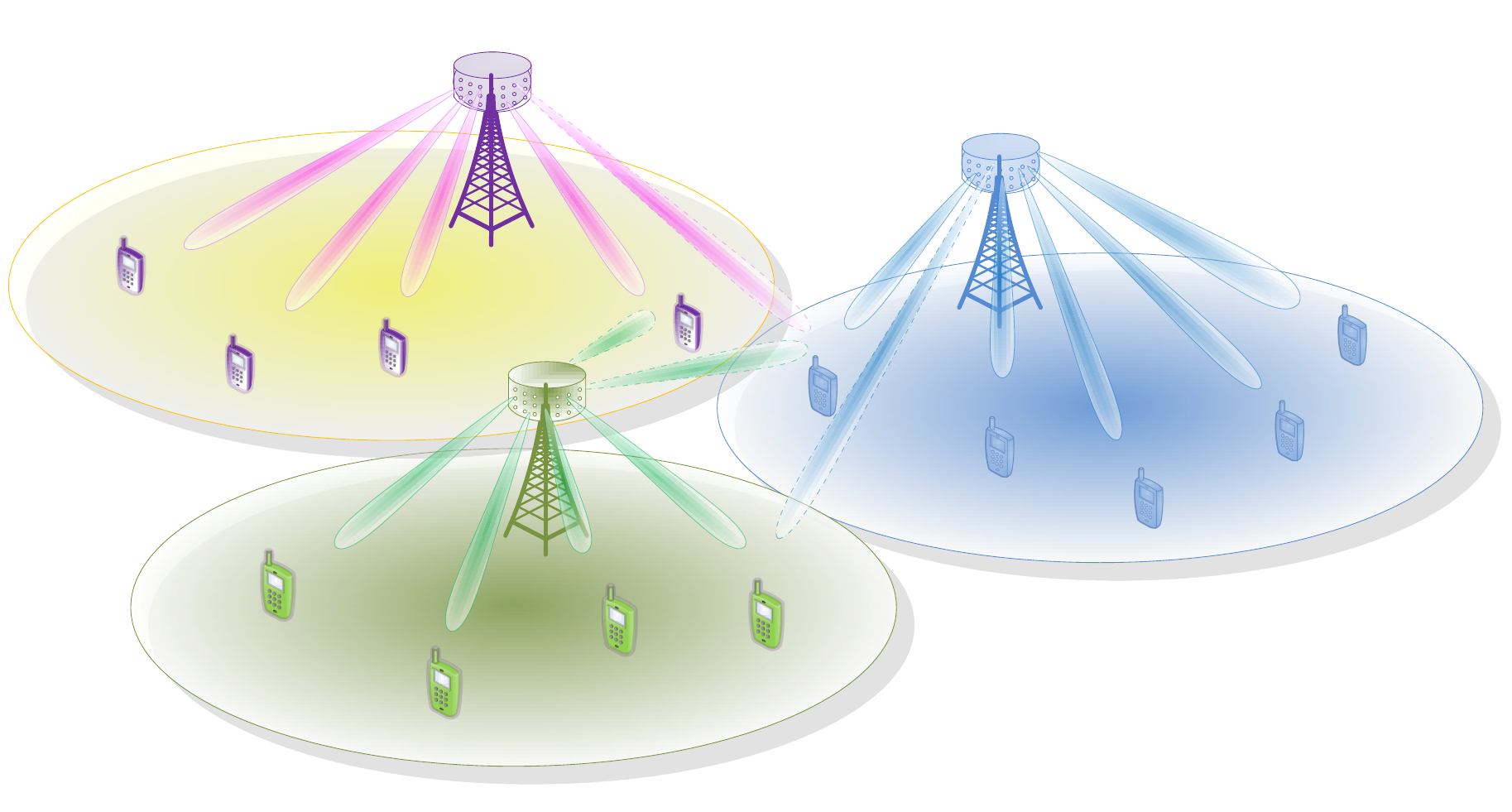}
\caption{System model of the multi-cell massive MIMO.}
\label{scenario}
\end{figure}

    Consider a multi-cell massive MIMO system consisting of \(L\) cells where each BS is equipped with an array of \(M\) antennas, and each cell is filled with \(K\) single-antenna UTs uniformly as illustrated in Fig. \ref{scenario}.
    Assume \(M \gg K\).
    The focus of this paper is on the uplink without any form of BS cooperation.
    Let \(g_{limk}\) denote the complex propagation coefficient between the \(m\)-th BS antenna in the \(l\)-th cell and the \(k\)-th UT in the \(i\)-th cell.
    Then, it can be expressed as,
\begin{equation}\begin{split}\label{channel gain}
{g_{limk}} = {h_{limk}}\sqrt{{\beta_{lik}}} \ ,
\end{split}\end{equation}
    where the small-scale fading coefficient \(h_{limk}\) is always assumed to be i.i.d. random variable with distribution \(\mathcal{CN}(0,1)\), and the large-scale fading coefficient \(\sqrt {{\beta _{lik}}}\) models the geometric attenuation and shadow fading, which is assumed to be independent over \(m\), constant over many coherence time intervals and known a priori \cite{Noncooperative}.
    The component \(\beta_{lik} = \varphi {\zeta_{lik}}/d_{lik}^\alpha\) consists of path loss and shadow fading, where \(\varphi\) is a constant related to carrier frequency and antenna gain, \(d_{lik}\) is the distance between the BS in the \(l\)-th cell and the \(k\)-th UT in the \(i\)-th cell, \(\alpha\) is the path loss exponent, and \(\zeta_{lik}\) represents the shadow fading with the distribution \(10{\log_{10}}{\zeta_{lik}}\sim \mathcal{N}(0,\sigma _{sh}^2)\).
    Then, we have propagation matrix \({\boldsymbol{G}}_{li} = {{\boldsymbol{H}}_{li}}{{\boldsymbol{D}}_{li}^{1/2}}\), where \({\boldsymbol{H}}_{li}\) denotes the \(M \times K\) matrix of fast fading coefficients between the BS in the \(l\)-th cell and the \(K\) UTs in the \(i\)-th cell, i.e., \({[{{\boldsymbol{H}}_{li}}]_{mk}} = {h_{limk}}\), and \({\boldsymbol{D}}_{li}\) is the \(K \times K\) diagonal matrix with \({[{{\boldsymbol{D}}_{li}}]_{kk}} = {\beta_{lik}}\).

    In the uplink, let \({{\boldsymbol{y}}_l}\) denote the \(M \times 1\) received signal vector of the BS in the \(l\)-th cell. Then, it can be expressed as:
\begin{equation}\begin{split}\label{received signal}
{{\boldsymbol{y}}_l} = \sum\limits_{i = 1}^L {{{\boldsymbol{G}}_{li}}{\boldsymbol{P}}_i^{1/2}{{\boldsymbol{x}}_i}} + {{\boldsymbol{n}}_l} \ ,
\end{split}\end{equation}
    where \({{\boldsymbol{x}}_i} \in {\mathbb{C}^{K \times 1}}\) denotes the transmit symbol vector in the \(i\)-th cell {with} \(\boldsymbol{x}_i \sim \mathcal{CN}(\bold{0},\bold{I}_K)\), 
    \({\boldsymbol{P}}_i^{1/2} = \text{diag}\{ \sqrt {{p_{i1}}} ,\sqrt {{p_{i2}}} , \cdots ,\sqrt {{p_{iK}}} \}\) denotes the transmit power matrix allocated to the UTs in the \(i\)-th cell, and \({{\boldsymbol{n}}_l} \in {\mathbb{C}^{M \times 1}}\) denotes the additive white Gaussian noise (AWGN) {with} \(\boldsymbol{n}_l \sim \mathcal{CN}(\bold{0},\sigma_n^2{\bold{I}_M)}\).

    We consider the case where the BSs have the perfect channel state information (CSI), i.e., they know \(\boldsymbol{G}\).
    Assume that the zero-forcing (ZF) receiver is utilized to detect the streams of the \(K\) UTs.
    Let \({\boldsymbol{A}}_{l}\) be the receiver matrix, \({\boldsymbol{a}}_{lk}\) the \(k\)-th column of \({\boldsymbol{A}}_{l}\), \({\boldsymbol{g}}_{llk}\) the \(k\)-th column of the propagation matrix \({\boldsymbol{G}}_{ll}\), and \({p_{lk}}\) the transmit power allocated to the \(k\)-th UT in the \(l\)-th cell.
    Then, the detected signal of the \(k\)-th UT in the \(l\)-th cell can be expressed as:
\begin{equation}\begin{split}\label{actual signal}
{z_{lk}} = {\boldsymbol{a}}_{lk}^H{{\boldsymbol{y}}_l} = & p_{lk}^{1/2}{\boldsymbol{a}}_{lk}^H{{\boldsymbol{g}}_{llk}}{x_{lk}} + {\boldsymbol{a}}_{lk}^H\sum\limits_{\kappa=1\hfill\atop \kappa\ne k}^K {p_{l\kappa }^{1/2}{{\boldsymbol{g}}_{ll\kappa }}{x_{l\kappa }}}\\  &+ {\boldsymbol{a}}_{lk}^H\sum\limits_{i=1\hfill\atop i\ne l}^L {{{\boldsymbol{G}}_{li}}{\boldsymbol{P}}_i^{1/2}{{\boldsymbol{x}}_i}}  + {\boldsymbol{a}}_{lk}^H{{\boldsymbol{n}}_l} \ ,
\end{split}\end{equation}
    where only the first term is the desired information, while the other terms represent the intra-cell interference, inter-cell interference and noise, respectively.
    As a result, the uplink signal to interference plus noise ratio (SINR) of the \(k\)-th UT in the \(l\)-th cell can be expressed as follows:
\begin{multline}\label{SINR}
{\gamma}_{lk}=\\
\frac{{{p_{lk}}{{\left| {{\boldsymbol{a}}_{lk}^H{\boldsymbol{g}}_{llk}^{}} \right|}^2}}}{{\sum\limits_{\kappa=1\hfill\atop \kappa\ne k\hfill}^K {p_{l\kappa }^{}{{\left| {{\boldsymbol{a}}_{lk}^H{\boldsymbol{g}}_{ll\kappa }^{}} \right|}^2}}  + \sum\limits_{i=1\hfill\atop i\ne l}^L {\sum\limits_{\kappa  = 1}^K {p_{i\kappa }^{}{{\left| {{\boldsymbol{a}}_{lk}^H{\boldsymbol{g}}_{li\kappa }^{}} \right|}^2}} }  + \sigma _n^2{{\left\| {{\boldsymbol{a}}_{lk}^{}} \right\|}^2}}} \ .
\end{multline}

    The EE of a communication system is measured in bits/Joule and defined as the total average number of bits/Joule successfully delivered from the UTs.
    As for the detailed power consumption model, for a specific UT, apart from the power consumed at the UT which can be modeled as the sum of transmit power and circuit power consumed by inevitable electronic operations, the average circuit power consumption within the BS is of great importance such as receiver antenna units, decoding, multiuser detection and fixed power consumption.
    For readers not interested in the receiver circuit power, the average circuit power consumption in BS can be assumed to be zero \cite{Power}.
    In this paper, our focus lies in the power allocation scheme, without consideration of other parameters such as \(M\) or \(K\), so there is no need to formulate such a trivial power consumption model as \cite{Impact}.
    Hence, we integrate all of the power consumed above but the transmit power into a specific term, \(p_{lc}\), denoted as the average circuit power consumption for each UT in the \(l\)-th cell, to simplify the power consumption model.
    Therefore, the uplink EE of the multi-cell massive MIMO system is given by
\begin{equation}\begin{split}\label{EE}
EE = \sum\limits_{l = 1}^L {\sum\limits_{k = 1}^K {\frac{{{{\log }_2}(1 + {\gamma}_{lk})}}{{{p_{lk}} + {p_{lc}}}}} } \ .
\end{split}\end{equation}	
	
\section{The Proposed CMDP-based Power Allocation Scheme}

    Using the system model presented in Section II, we formulate the power allocation optimization problem by applying the CMDP, and then propose an offline solution containing the value iteration and Q-learning algorithms to solve it.

\subsection{Formulation of Optimization Problem}

    We first extract the main characters from the above system model to build a CMDP-based model.
    A CMDP-based model can be characterized by five elements: decision epochs, states, actions, transition probabilities and rewards \cite{Book}.
	
    \({\textsl{Decision Epochs:}} \) Before modeling, the time dimension is partitioned into decision slots represented by \(\{ 1,2, \cdots ,n, \cdots \}\), where the time slot \(n\) is defined as the time interval \(\left[ {nT_c,(n + 1)T_c} \right]\), and \(T_c\) denotes the channel coherence time.
    Then, the decision epochs can be indexed with \(n\).
    We assume that the wireless channel fluctuates slowly and the CSI remains quasi-static and i.i.d. between decision slots.

    \({\textsl{States:}}  \) To model the fluctuation in the physical layer, a finite-state Markov channel (FSMC) model can be built to characterize the time-varying behavior of the channel \cite{FSMC}.
    In our model, the system state space \({S^C} = {C^S} \times {C^S} \times  \cdots  \times {C^S}\) is the Cartesian product of cell state space \({C^S}\) accounting for the channel gains in each cell, whose component is also a composite state of link state \({\boldsymbol{g}}_{llk}^H{{\boldsymbol{g}}_{li\kappa}}\), denoted by \({\psi _{lik\kappa}}\), and each link state is quantized using a finite number of thresholds \(\Gamma  = \{ 0 = {\Gamma _0},{\Gamma _1}, \cdots ,{\Gamma _{{Q_S}}} = \infty \} \), where \({\Gamma _b} < {\Gamma _{b'}}, \ \forall \  b < b'\).
    The composite system state of all the cells is denoted by \({s_c} = \{ {s_1},{s_2}, \cdots ,{s_L}\} \), where \({s_l} = \{ {\psi_{lik\kappa}}|i = 1, \cdots ,L; \ k,\kappa = 1, \cdots ,K\} \).
    Based on the above assumptions, the sequence of composite system states forms a Markov chain with transition probabilities \(P \{ {s_c^{'}}|{s_c} \} \) that is independent of actions, which is similar as that in \cite{SMDP}, and \({s_c}\), \({s_c^{'}}\) denote the system states in current and next decision epoch, respectively.

    \({\textsl{Actions:}}  \) Let \({A^C} = {C^A} \times {C^A} \times  \cdots  \times {C^A}\) denote the system action space, and \(C^A\) denote the action space of each cell, whose cardinality is \({Q_A}\).
    Specially, let \(a_{p,l}\) denote {the set of the transmit powers allocated to the UTs in the \(l\)-th cell.}
    Then, the composite system action can be denoted by \({a_c} = \{ {a_{p,1}},{a_{p,2}},\cdot\cdot\cdot,{a_{p,L}}\} \).
    Note that the proper choice of the set of action space can incorporate the QoS requirement with respect to the maximum transmit power for each UT without additional operations.

   \({\textsl{Transition \ Probabilities:}}  \) Based on the modeling of FSMC, the link state transition occurs only from the current state to its neighboring states.
   Without loss of generality, we simplify each link state \({\psi _{lik\kappa}}\) as \(\psi\) for convenience.
   Then, the steady probability for the \(b\)-th link state can be expressed as
\begin{equation}\begin{split}\label{steady probability}
p_b = \int_{{\Gamma _b}}^{{\Gamma _{b + 1}}} {\frac{1}{{{\psi _0}}}{e^{ - \frac{\psi }{{{\psi _0}}}}}d\psi }  = {e^{ - {\textstyle{{{\Gamma _b}} \over {{\psi _0}}}}}} - {e^{ - {\textstyle{{{\Gamma _{b + 1}}} \over {{\psi _0}}}}}} \ ,
\end{split}\end{equation}
    where \({\psi _0} = \mathbb{E}\{ \psi \} \) is the average link gain.
    According to \cite{FSMC}, the level-crossing rate of the link gain is given by \(h(\psi ) = \sqrt {2\pi \psi /{\psi _0}} {f_c}{e^{ - {\textstyle{\psi  \over {{\psi _0}}}}}}\), where \({f_c}\) is the maximum Doppler frequency normalized by the decision rate \(1/T_c\).
    The link state transition probabilities are determined by
\begin{multline}\label{transition probability}
\begin{array}{l}
p\{ {\psi ^{'}} = b'|{\psi} = b\}\\[0.5pt] \\
 = \left\{ {\begin{array}{*{20}{l}}
{\frac{{h({\Gamma _{b + 1}})}}{{{p_b}}},  b' = b + 1,  b \in [0,{Q_S} - 2]  ,}\\[0.5pt] \\
{\frac{{h({\Gamma _b})}}{{{p_b}}},  b' = b - 1,  b \in [1,{Q_S} - 1]  ,}\\[0.5pt] \\
{1 - \frac{{h({\Gamma _{b + 1}})}}{{{p_b}}} - \frac{{h({\Gamma _b})}}{{{p_b}}},  b' = b,  b \in [1,{Q_S} - 2] ,}
\end{array}} \right.
\end{array}
\end{multline}
    where \({\psi}\) and \({\psi ^{'}}\) denote the link states in current and next decision epoch, respectively.
    The transition probabilities of \(p\{ {\psi ^{'}} = {b'}|{\psi } = b\}\) for the boundaries are given by \(p\{ {\psi ^{'}} = 0|{\psi } = 0\}  = 1 - p\{ {\psi ^{'}} = 1|{\psi } = 0\}\) and \(p\{ {\psi ^{'}} = {Q_S} - 1|{\psi } = {Q_S} - 1\} = 1 - p\{ {\psi ^{'}} = {Q_S} - 2|{\psi } = {Q_S} - 1\}\).

    The composite system state transition probabilities can be computed by
\begin{equation}\begin{split}\label{transition probability_final}
P \{ s^{'}_c|s_c,a_c \} = \prod\limits_{l = 1}^L {\prod\limits_{i = 1}^L {\prod\limits_{k = 1}^K {\prod\limits_{\kappa = 1}^K {p\{ {\psi ^{'}_{lik\kappa}}|\psi _{{lik\kappa}}\} } } } } \ .
\end{split}\end{equation}

    \({\textsl{Rewards:}} \) We adopt the overall EE as system reward function, which is defined as \(R({s_c},{a_c})\) for the action \({a_c}\) at the state \({s_c}\).
    And the corresponding QoS requirement with respect to the minimum data rate for each UT can be expressed as a series of constraints: \({C_{lk}}({s_c},{a_c}) \ge {r_{\min }}\), for \(l = 1, \cdots ,L\) and \(k = 1, \cdots ,K\), where \(C_{lk}\) and \(r_{\min}\) denote the instantaneous data rate and the required minimum data rate for each UT in the uplink, respectively.

    By exploiting the above CMDP framework, the transmit power can be adjusted according to a stationary policy \(\pi  = ({\delta _1},{\delta _2}, \cdots ,{\delta _n}, \cdots )\), where each decision rule \({\delta _n}\) specifies a mapping function \({\delta _n}:{S^C} \to {A^C}\) to maximize the objective function.
    Let \(\lambda\) denote the discount factor, \({v^\pi }(s^0_c)\) denote the expected discounted total reward, and \({c_{lk}^\pi }(s^0_c)\) denote the expected discounted total cost associated with the required date rate constraint, given that the policy \(\pi\) is used with initial state \(s^0_c\).
    Then, we can formulate the CMDP-based optimization problem as follows
\begin{equation}\begin{split}\label{OP}
\mathop {\max }\limits_{a^n_c} \mathop {}\nolimits^{} \quad & {v^\pi }(s^0_c) = {\rm \mathbb{E}}_{{s^0_c}}^\pi \left\{ \sum\limits_{n = 1}^\infty  {{\lambda ^{n - 1}}R(s^n_c,a^n_c)} \right\} \\
\text{s.t.  } \quad & { c_{lk}^\pi }(s^0_c) = {\mathbb E}_{s^0_c}^\pi \left\{ \sum\limits_{n = 1}^\infty  {{\lambda ^{n - 1}}{C_{lk}}(s^n_c,a^n_c)} \right\}  \ge {r_{\min }} \ , \\
& l = 1, \cdots, L, \ k = 1, \cdots, K.
\end{split}\end{equation}

\subsection{Offline Solution}

    To solve the constrained optimization problem in (\ref{OP}), we first utilize the Lagrangian approach \cite{Book}, \cite{Q-learning} to transform the CMDP optimization problem into an equivalent unconstrained MDP optimization problem.
    For any non-negative vector of Lagrange multipliers (LM) \({\boldsymbol{\rho}} = {[\, {\rho _{lk}} \, | \, l = 1, \cdots, L, \, k = 1, \cdots, K \,]^T}\), we define the Lagrangian as
\begin{equation}\begin{split}\label{Lagrangian}
L(s_c,a_c;{\boldsymbol{\rho}}) = R(s_c,a_c) + \sum\limits_{l=1,k=1}^{L,K}{{\rho_{lk}}{C_{lk}}(s_c,a_c)} \ .
\end{split}\end{equation}
	Then, the Bellman's equations are given by
\begin{multline}\label{Bellman}
{v_{\boldsymbol{\rho }}}(s_c) = \\
 \mathop {\max }\limits_{a_c} \left\{ {L(s_c,a_c;{\boldsymbol{\rho }}) + \sum\limits_{s_c^{'} } {\lambda P\{ s_c^{'}|s_c,a_c\} {v_{\boldsymbol{\rho }}}(s_c^{'})} } \right\}  .
\end{multline}

    Now we propose an offline scheme to derive the optimal power allocation policy.
    The stationary optimal policy and the corresponding maximum expected discounted total reward function can be obtained by the well-known value iteration algorithm \cite{Book}, for a fixed LM vector \({\boldsymbol{\rho}}\).
    Then, we utilize the Q-learning algorithm \cite{Q-learning} to determine the proper \({\boldsymbol{\rho}}\) for the feasible constraint \({r_{\min }}\).
    Specifically, the iteration algorithm is described as follows
\begin{equation}\begin{split}\label{Q}
{\rho _{lk,j' + 1}} = {\rho _{lk,j'}} + \frac{1}{j'}\left( {{r_{\min }} - c_{lk}^{{\pi}^*} (s_c^0)} \right),
\end{split}\end{equation}
    where \(j'\) is the index of the iteration steps.
    The convergence to the global optimum \({\boldsymbol{\rho}}^*\) of the Q-learning algorithm can be ensured, because the functions
\begin{multline}\label{conv}
\int_0^{{\rho _{lk}}} {\left( {{r_{\min }} - c_{lk}^{{\pi ^*}}(s_c^0)} \right)d{\rho _{lk}}}, \\
 l=1, \cdots, L, \ k=1, \cdots, K
\end{multline}
    are piece-wise linear concave\cite{Q-learning}.
    Taking into consideration of the convergence to the global optimum policy \({\pi}^*\) ensured by the value iteration algorithm \cite{Book}, the proposed offline algorithm can attain the global optimum power allocation scheme.

    The offline iterative algorithm is summarized in Algorithm 1, where \(\epsilon\) denotes an infinitesimal gap, and \(i'\) is the index of the iteration steps for the value iteration algorithm.
    We remark here that the inner iteration between step 2 and step 5 in Algorithm 1 performs the value iteration computation to solve the Bellman's equations in (\ref{Bellman}) to obtain the stationary optimal policy for the given LM vector \({\boldsymbol{\rho }}_{j'}\), where we denote the \(i'\)-th approximation to \({v_{\boldsymbol{\rho }}}(\cdot)\) by \({v_{i'}^{\pi}}(\cdot)\).
    We also remark here that the outer iteration between step 2 and step 5 in Algorithm 1 performs the Q-learning computation to solve the equations in (\ref{Q}) to obtain the optimal LM vector \({\boldsymbol{\rho}}^*\), where we replace \(c_{lk}^{{\pi}^*} (s_c^0)\) by \(c_{lk}^{{\pi}^*({{\boldsymbol{\rho }}_{j'}})} (s_c^0)\) in the \(j'\)-th iteration.
    We point out here that the value of the components of the initial LM vector \({\boldsymbol{\rho }}_0\) should be set to be very large to converge to the optimal LM vector in consideration of the minimum data rate constraint.

    The obtained optimal decision policy \({\pi}^*({{\boldsymbol{\rho }}^*})\) in Algorithm 1 contains a series of optimal decision rule \({\delta ^*}\), which specifies a mapping function \({\delta ^*}:{S^C} \to {A^C}\) to get the maximum reward.
    The mapping function \({\delta ^*}\) can be exploited to construct an offline look-up table to avoid the frequent and continuous computations.
    By using the table, the corresponding transmit power can be allocated to the UTs to maximize the reward once the channel states are known.
    Note here that tradeoff between the performance gain and the size of the offline table can be balanced by changing \({Q_S}\) and \({Q_A}\).
\begin{algorithm}[!tbp]
\doublespacing
\caption{Offline Algorithm to Solve the CMDP-based Optimization Problem}
\begin{itemize}
\item Step 1: \ Set \({v_0^{\pi}}(s_c)=0\) for each \({s_c}\). Specify \({\epsilon}>0\), \({\boldsymbol{\rho }}_0>\bf{0}\), and initialize \(i'=0\), \(j'=0\).
\item Step 2: \ For a given \({\boldsymbol{\rho }}_{j'}\), compute \({v_{{i'}+1}^{\pi}}(s_c)\) for each \({s_c}\) as follows
\begin{multline*}
{v_{{i'}+1}^{\pi}}(s_c)= \\
\mathop {\max }\limits_{a_c} \left\{ {L(s_c,a_c;{\boldsymbol{\rho }}_{j'})+\sum\limits_{s_c^{'} } {\lambda P\{ s_c^{'}|s_c,a_c\} {v_{i'}^{\pi}}(s_c^{'})} } \right\}.
\end{multline*}
\item Step 3: \ If \(\left\| {v_{i'+1}^{\pi}}-{v_{i'}^{\pi}} \right\| < \epsilon (1 - \lambda )/\lambda\), then \({\pi}^*({{\boldsymbol{\rho }}_{j'}})= {\pi}\) and continue. Otherwise, increase \(i'\) by 1 and goto step 2.
\item Step 4: \ For each \(l\) and \(k\), compute \({\rho _{lk,j' + 1}}\) as follows\\
\({\rho _{lk,j' + 1}} = {\rho _{lk,j'}} + \frac{1}{j'}\left( {{r_{\min }} - c_{lk}^{{\pi}^*({{\boldsymbol{\rho }}_{j'}})} (s_c^0)} \right)\).
\item Step 5: \ If \(\left\| {{\boldsymbol{\rho }}_{j'+1}}-{{\boldsymbol{\rho }}_{j'}} \right\| < \epsilon\), then \({\boldsymbol{\rho }}^* = {\boldsymbol{\rho }}_{j'}\) and continue. Otherwise, increase \(j'\) by 1 and goto step 2.
\item Step 6: \ Compute the optimal policy \({\pi}^*({{\boldsymbol{\rho }}^*})\) according to step 2 and step 3, with the optimal LM factor \({\boldsymbol{\rho }}^*\).
\end{itemize}
\label{Algorithm}
\end{algorithm}

\section{Simulation Results}

    In this section, we compare the performances of our proposed CMDP-based power allocation policy with the ergodic optimal policy.
    The ergodic optimal policy is achieved by maximizing the reward functions from the set of feasible actions at each system state by using the exhaustive-search method.
    Note that the focus of the comparison between our proposed policy and the ergodic optimal policy does not lie in the uniformity, i.e., specific action chosen at each system state, but the long-term performance with respect to the expected discounted total reward.
    In our simulations, unless otherwise stated, the system parameters are set as follows: the number of cells is \({L=2}\), the number of UTs in each cell is \({K=1}\), the path loss factor \(\varphi\) is 1, the path loss exponent \(\alpha\) is 3.7, the variance of log-normal shadow fading \(\sigma _{sh}^2\) is 10dB, the number of antennas to each BS is \(M=128\), the average circuit power consumption is \({p_{lc}}=10 \text{mw}\) for each \({l}\) \cite{Impact}, the link states are equiprobable, quantized with \({Q_S} = 4\) states, the action at each system state is chosen from \({\rm{{10^{-2}}mw}}\) to \({\rm{{10^{2}}mw}}\) with \({Q_A} = 20\) intervals, and the discount factor \(\lambda\) is assumed to be \({0.9}\).

\begin{figure}[t]
\centering
\includegraphics[width=0.45\textwidth]{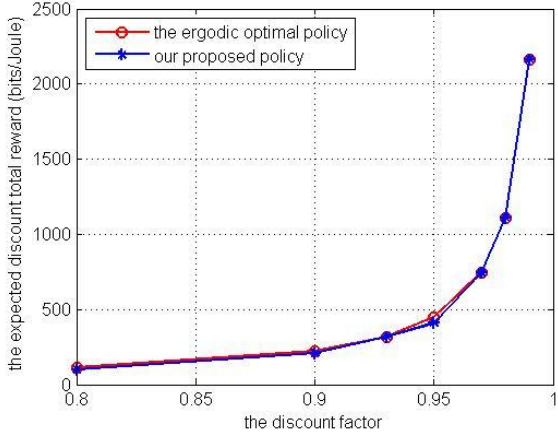}
\caption{The expected discounted total reward vs. the discount factor.}
\label{Simulation_comparison}
\end{figure}

    Fig. \ref{Simulation_comparison} shows the performances of our proposed policy in comparison with the ergodic optimal policy.
    It is clear that the performance gap between them can be negligible.
    And the expected discounted total reward under different discount factors increases as \({\lambda}\) becomes larger.
    This is because larger discount factor means longer-term reward taken into consideration.

\begin{figure}[!htbp]
\centering
\includegraphics[width=0.45\textwidth]{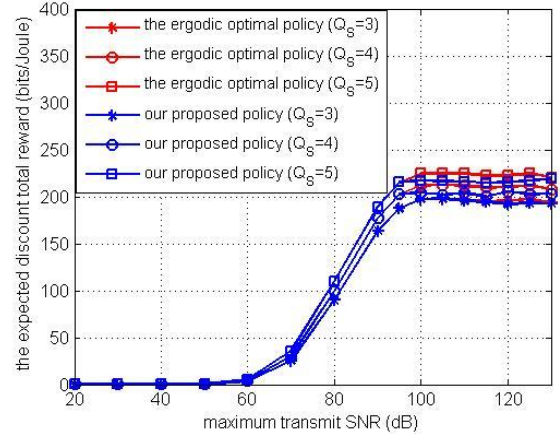}
\caption{The expected discounted total reward vs. the maximum transmit SNR allocated to the UTs with different \({Q_S}\) under \({Q_A=20}\).}
\label{Simulation_max_snr_QS}
\end{figure}

\begin{figure}[!htbp]
\centering
\includegraphics[width=0.45\textwidth]{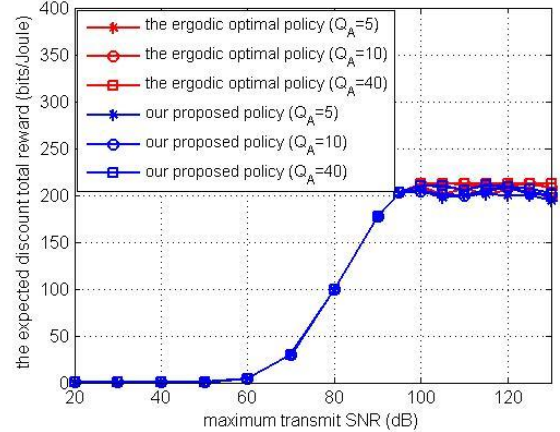}
\caption{The expected discounted total reward vs. the maximum transmit SNR allocated to the UTs with different \({Q_A}\) under \({Q_S=3}\).}
\label{Simulation_max_snr_QA}
\end{figure}

    Fig. \ref{Simulation_max_snr_QS} and Fig. \ref{Simulation_max_snr_QA} show how the maximum transmit signal-to-noise ratio (SNR) allocated to the UTs affects the expected discounted total reward of policies with \(\sigma_n^2 = -101\text{dBm}\).
    It can be observed that the performances of the two policies stop to increase and tend to be constant when the maximum transmit SNR is larger than certain threshold SNR.
    The reason is that there is no longer any need to consume more power, when the maximum expected total reward has already been achieved.
    In Fig. \ref{Simulation_max_snr_QS}, it can be observed that the performances of both policies improve as \({Q_S}\) gets larger, which is due to the more refined quantification of the channel states.
    {It can also be observed that there is a performance gap between the two policies when the maximum transmit SNR is large enough.
    This results from that the iteration process of the value iteration algorithm makes our proposed policy  sensitive to the channel state quantized by \({Q_S}\), which will be aggravated by the larger maximum transmit SNR.}
    In Fig. \ref{Simulation_max_snr_QA}, it can be observed that the performances of the two policies improve as \({Q_A}\) gets larger when the maximum transmit SNR is large enough.
    {The reason is that the larger \({Q_A}\), the higher-precision transmit power can be allocated to the UTs.}
    In addition, we can see that the performance gain by increasing \({Q_A}\) is no longer significant when \({Q_A}\) gets large enough.
    This reveals that the proposed policy can achieve good performance with a small \({Q_A}\), i.e., a low-precision transmit power quantization is enough for our proposed policy.

\begin{figure}[!htbp]
\centering
\includegraphics[width=0.45\textwidth]{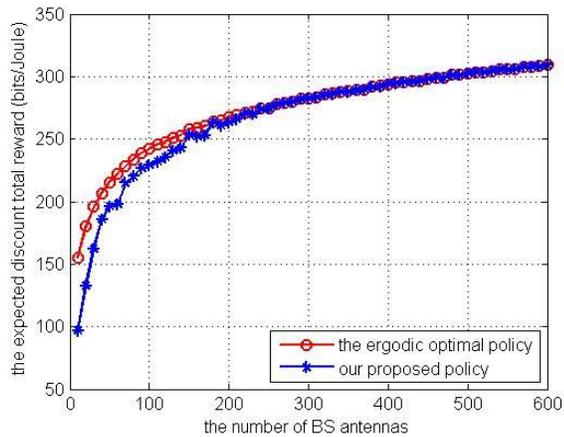}
\caption{The expected discounted total reward vs. the number of BS antennas.}
\label{Simulation_M}
\end{figure}

    In Fig. \ref{Simulation_M}, the impact of the number of antennas in BS on the performance of the policies is shown.
    We can see that the performance growth tends to slow down with the increase of \(M\).
    The performance of our proposed policy is very close to the ergodic optimal policy when \(M\) is large enough, but not for the case of small \(M\).
    Recall that the value iteration algorithm is sensitive to the channel state based on \({\boldsymbol{g}}_{llk}^H{{\boldsymbol{g}}_{li\kappa}}\) which contains the parameter \(M\), and that the large-scale fading coefficient \(g_{limk}\) is independent of \(M\).
    {Correspondingly, for a smaller \(M\), the impact of the small-scale fading coefficient of the channel will be increased sharply which results in the worse modeling of the channel state.}
    This leads to  the performance gap of the two policies for the case of smaller \(M\).

\section{Conclusions}

    In this paper, we have proposed a CMDP-based power allocation algorithm for the uplink of multi-cell massive MIMO system to maximize EE under two QoS requirements. The policy performance of our CMDP-based offline algorithm is very close to the ergodic optimal policy, and some further analysis results have been given to determine the impact of system parameters on the long-term EE.

\section*{Acknowledgments}

    This work was supported in part by the National Basic Research Program of China (973 Program 2012CB316004), the National 863 Project (2015AA01A709), and the Natural Science Foundation of China (61221002).

\bibliographystyle{IEEEtran}

\begin{thebibliography}{10}
\providecommand{\url}[1]{#1}
\csname url@samestyle\endcsname
\providecommand{\newblock}{\relax}
\providecommand{\bibinfo}[2]{#2}
\providecommand{\BIBentrySTDinterwordspacing}{\spaceskip=0pt\relax}
\providecommand{\BIBentryALTinterwordstretchfactor}{4}
\providecommand{\BIBentryALTinterwordspacing}{\spaceskip=\fontdimen2\font plus
\BIBentryALTinterwordstretchfactor\fontdimen3\font minus
  \fontdimen4\font\relax}
\providecommand{\BIBforeignlanguage}[2]{{%
\expandafter\ifx\csname l@#1\endcsname\relax
\typeout{** WARNING: IEEEtran.bst: No hyphenation pattern has been}%
\typeout{** loaded for the language `#1'. Using the pattern for}%
\typeout{** the default language instead.}%
\else
\language=\csname l@#1\endcsname
\fi
#2}}
\providecommand{\BIBdecl}{\relax}
\BIBdecl

\bibitem{Noncooperative}
T.~L. Marzetta, ``Noncooperative cellular wireless with unlimited numbers of
  base station antennas,'' \emph{IEEE Trans. Wireless Commun.}, vol.~9, no.~11,
  pp. 3590--3600, Nov. 2010.

\bibitem{Optimal}
E.~Bjrnson, L.~Sanguinetti, J.~Hoydis, and et~al., ``Optimal design of
  energy-efficient multi-user {MIMO} systems: is massive {MIMO} the answer?''
  \emph{IEEE Trans. Wireless Commun.}, vol.~14, no.~6, pp. 3059--3075, June
  2015.

\bibitem{Tradeoff}
S.~Mukherjee and S.~K. Mohammed, ``On the energy-spectral efficiency trade-off
  of the {MRC} receiver in massive {MIMO} systems with transceiver power
  consumption,'' \emph{arXiv:1404.3010v1 [cs.IT]}, Apr. 2014.

\bibitem{Impact}
S.~K. Mohammed, ``Impact of transceiver power consumption on the energy
  efficiency of zero-forcing detector in massive {MIMO} systems,'' \emph{IEEE
  Trans. Commun.}, vol.~62, no.~11, pp. 3874--3890, Nov. 2014.

\bibitem{SMDP}
P.~Wang, X.~Zhang, and M.~Song, ``Optimal stochastic subcarrier and power
  allocations for {QoS}-guaranteed services in {OFDMA} multicell cooperation
  networks,'' \emph{in Proc. IEEE ICC}, pp. 6449--6453, June 2013.

\bibitem{MIMO}
D.~V. Djonin and V.~Krishnamurthy, ``{MIMO} transmission control in fading
  channels -- a constrained {Markov} decision process formulation with monotone
  randomized policies,'' \emph{IEEE Trans. Signal Process.}, vol.~55, no.~10,
  pp. 5069--5083, Oct. 2007.

\bibitem{OFDM}
K.~Bi, Q.~Yang, F.~Fu, and et~al., ``Energy-efficient power and subcarrier
  allocation for {OFDMA} systems with value function approximation approach,''
  \emph{in Proc. IEEE ICTC}, pp. 530--535, Oct. 2012.

\bibitem{Power}
G.~Miao, ``Energy-efficient uplink multi-user {MIMO},'' \emph{IEEE Trans.
  Wireless Commun.}, vol.~12, no.~5, pp. 2302--2313, May 2013.

\bibitem{Book}
E.~Altman, \emph{Constrained {Markov} decision processes: stochastic
  modeling}.\hskip 1em plus 0.5em minus 0.4em\relax London: Chapman and Hall
  CRC, 1999.

\bibitem{FSMC}
Q.~Zhang and S.~A. Kassam, ``Finite-state {Markov} model for {Rayleigh} fading
  channels,'' \emph{IEEE Trans. Commun.}, vol.~47, no.~11, pp. 1688--1692, Nov.
  1999.

\bibitem{Q-learning}
D.~V. Djonin and V.~Krishnamurthy, ``Q-learning algorithms for constrained
  {Markov} decision processes with randomized monotone policies: application to
  {MIMO} transmission control,'' \emph{IEEE Trans. Signal Process.}, vol.~55,
  no.~5, pp. 2170--2181, May 2007.

\end{thebibliography}


\end{document}